\documentclass[usenatbib]{mn2e}
\usepackage{times}
\usepackage{epsfig}
\usepackage{amsmath}

\def \aj {AJ}
\def \mnras {MNRAS}
\def \apj {ApJ}
\def \apjs {ApJS}
\def \apjl {ApJL}
\def \aap {A\&A}
\def \nat {Nature}
\def \pasp {PASP}
\def \araa {ARA\&A}

\title[SFR at $z\sim1$]{Dependence of Star Formation Activity On Stellar Mass and Environment From the Redshift One LDSS-3 Emission Line Survey (ROLES)}
\author[I.H. Li et al]{I.H. Li$^{1,*}$, Karl Glazebrook$^{1}$, David Gilbank$^{2}$, Michael Balogh$^{2}$, Richard Bower$^{3}$, 
\newauthor Ivan Baldry$^{4}$, Greg Davies$^{3}$, George Hau$^{1}$ and Pat McCarthy$^{5}$\\
$^{1}${Centre for Astrophysics \& Supercomputing, Swinburne University of Technology, PO Box 218, Hawthorn, Victoria 3122, Australia}\\
$^{2}${Department of Physics and Astronomy, University of Waterloo, Waterloo, Ontario, Canada, N2L 3G1}\\
$^{3}${Institute for Computational Cosmology, Department of Physics, University of Durham, South Road, Durham, DH1 3LE, UK}\\
$^{4}${Astrophysics Research Institute, Liverpool John Moores University, Twelve Quays House, Egerton Wharf, Birkenhead Ch41 1LD, UK}\\
$^{5}${Carnegie Observatories, 813 Santa Barbara Street, Pasadena, California, 91101 USA}\\
$^{*}${email:tli@astro.swin.edu.au}\\
}
\begin{document}
\maketitle
\begin{abstract}
Using the sample from the \it Redshift One LDSS3 Emission line Survey \rm (ROLES),
we probe the dependence of star formation rate (SFR) and specific star
formation rate (sSFR) as a function of stellar mass $M_*$ and environment as
defined by local galaxy density, in the CDFS field. Our spectroscopic sample consists of 312
galaxies with $K_{AB}<24$, corresponding to stellar mass
$\log(M_*/M_{\sun})>8.5$, and with [OII] derived star-formation rates
SFR$>0.3M_{\sun}/$yr, at $0.889\leq z \leq 1.149$.
The results have been compared directly with the Sloan Digital Sky
Survey Stripe 82 sample at $0.032\leq z \leq 0.05$. 
For star-forming galaxies, we confirm that there is little correlation between SFR and density at
$z\sim 0$.  However, for the lowest
mass galaxies in our $z\sim 1$ sample, those with $\log(M_*/M_{\sun})<10$, we find that both the median SFR and
specific SFR {\it increase} significantly with increasing local density.  The
``downsizing'' trend for low mass galaxies to be quenched progressively
later in time appears to be more pronounced in moderately overdense environments.  
Overall we find that the evolution of star-formation in galaxies is
most strongly driven by their stellar mass, with local galaxy density
playing a role that becomes increasingly important for lower mass galaxies.

\end{abstract}

\section{Introduction \label{introduction}}
The observed correlation between stellar mass and star formation rate
(SFR) in galaxies as a function of redshift provides insight into the
integrated SFR over the history of the Universe \citep[e.g.][]{2007ApJ...663..834B,2007ApJ...660L..43N}.
Several independent observations now suggest that the SFR in galaxies
is regulated and in some cases quenched by different physical processes
\citep[e.g.][]{2010arXiv1003.4747P}.  One of the fundamental questions
is whether these processes are more closely correlated with internal
galaxy properties (such as mass and luminosity) or with their external
environment.  

In particular, many studies have shown that the SFR in individual galaxies is affected by environment. 
In the local Universe, it is well established that the fraction of passive (red) galaxies is higher in regions of high galaxy density, while regions of low galaxy density harbor predominantly blue, star forming galaxies \citep[e.g.,][]{1980ApJ...236..351D,2004MNRAS.348.1355B,2006MNRAS.373..469B,2009ApJ...701..787P,2009MNRAS.393.1324B}.  This is also, perhaps more directly, reflected in the star formation rates of the galaxies 
\citep[e.g.,][]{2002MNRAS.334..673L,2003ApJ...584..210G,2004ApJ...601L..29H}.  Interestingly, it is primarily the fraction of star forming galaxies that correlates with environment \citep[e.g.,][]{2006MNRAS.373..469B,2009ApJ...698...83L,2010A&A...509A..40I,2010arXiv1003.4747P}; the distribution of colours or SFR among the active population itself seems surprisingly independent of environment, at least at low redshift \citep[e.g.,][]{2004ApJ...615L.101B,2005ApJ...629L..77Y,2001ApJ...559..606C,2005AJ....130.1482R,2007ApJS..172..270C,2009MNRAS.398..754B}.  
This may suggest that any environmental influence responsible for quenching SFR must operate on a short time scale of few Gyrs \cite[e.g.,][]{2004MNRAS.348.1355B,2010MNRAS.404.1231V}.
 
It is also well established that the global SFR density ($\rho_{SFR}$) has decreased steadily from $z\sim1$ to $z\sim0$ \citep[e.g.,][]{1996ApJ...460L...1L,1996MNRAS.283.1388M,2004ApJ...615..209H}.  The contribution of galaxies to $\rho_{SFR}$ depends on their stellar mass, with massive galaxies at higher-redshifts contributing a greater total  $\rho_{SFR}$ than similar mass galaxies today. This picture has been labeled `downsizing' --- the term denotes the general empirical observation that star-formation activity progresses from higher mass to lower mass systems with cosmic time  \citep{1996AJ....112..839C}.
There are various manifestations of this effect.  For example,  \cite{2005ApJ...619L.135J} and  \cite{2009ApJ...690.1074M} found that massive galaxies contributed a greater proportion of $\rho_{SFR}$ at high redshifts.  
Recently many surveys such as  GOODS
\citep[e.g.,][]{2009A&A...504..751S,2009A&A...496...57M}, GDDS
\citep[e.g.,][]{2004AJ....127.2455A,2005ApJ...619L.135J}, K20
\citep[e.g.,][]{2002A&A...381L..68C,2002A&A...384L...1D}, DEEP
\citep[e.g.,][]{2005ApJS..159...41V}, and zCOSMOS
\citep[e.g.,][]{2007ApJS..172...70L,2009arXiv0907.5416P,2010arXiv1007.3841C} have confirmed that stellar mass
plays a fundamental role in determining the fate of a galaxy.  

With large, spectroscopic surveys, several of these studies have begun
to explore the effect of environment over a wide range of redshifts
\citep[e.g.][]{2008MNRAS.383.1058C,2010arXiv1007.1967C,2009ApJ...694.1099M,2010arXiv1003.4747P,2010A&A...509A..40I,2010ApJ...710L...1V,2010arXiv1007.3841C},
and they generally find
that environment is a second-order effect compared with the mass-driven
evolution.  \cite{2010arXiv1003.4747P} has proposed a particularly
interesting, empirical model in which environment induces
transformation from active to passive, at a rate that is surprisingly
independent of stellar mass.  
  
Since environmental influences might be expected to have the largest
effect on gas-rich, low-mass galaxies, the relevant physics may thus
best be probed by studying such galaxies at $z=1$ or higher.  This is a
particularly interesting epoch, as there is mounting evidence that star
formation may actually be {\it enhanced} in gas--rich galaxies that
live in dense regions at these redshifts
\citep[e.g.,][]{2007A&A...468...33E,2008ApJ...686..966M,2008MNRAS.383.1058C,2009ApJ...700..971I,2010arXiv1007.2642S},
although the fraction of red (passively--evolving) galaxies still
increases with density \citep[e.g.][]{2010arXiv1007.1967C}.
However, most of the results cited above are based on flux-limited
surveys, and hence can only
spectroscopically probe the high-mass regime ($>10^{10}M_{\sun}$) at
$z\sim 1$.  For a complete picture it is
necessary to probe SFR in low-mass galaxies as well.

We have therefore carried out a spectroscopic survey, the `Redshift One LDSS3 Emission line Survey' (ROLES), to study the dependence of $\rho_{SFR}$ on stellar mass at $z\sim1$.
Our observations focus on low-mass galaxies
(8.5$\leq$log($M_*/M_{\sun}$)$\leq$10), $\sim$10 times less massive
than those typically considered in the current literature.
We aim to obtain a sample complete in both SFR and stellar mass.
The first two papers of this series, \citet{2009MNRAS.395L..76D} and \citet{2010MNRAS.405.2419G} (hereafter Paper I and Paper II), revealed that the $\rho_{SFR}$ does not increase continuously with decreasing stellar mass, but instead exhibits a `turn-over' at $\sim10^{10}M_{\sun}$ just as it does in the local Universe. 
This demonstrates that low mass galaxies do not dominate the total SFR density at any epoch $z\leq 1$.
Comparing the results with those derived from SDSS galaxies at $z\sim0$
using an identical [OII] calibration, Paper II finds that the shape of
the total SFR density as a function of stellar mass remains similar at
$z\sim1$ and $z\sim0$, although the magnitude of the total SFR density
becomes smaller at $z\sim0$. This implies that the rate at which the
SFR density decreases is not a strong function of stellar mass.
In other words, the evolution of global $\rho_{SFR}$ is better described as a simple evolution of the normalization of $\rho_{SFR}(M_*)$, 
and the shape and relative contributions of galaxies of different masses remained unchanged. 

In this paper, we address the star-forming population of galaxies at $z\sim$1, and use ROLES to probe how SFR correlates with environment at the lowest stellar masses considered to date.
The structure of the paper is as follows.
We present our survey data and catalogues in $\S$\ref{survey}.  
The environmental measure, local galaxy density, is presented in
$\S$\ref{glsd}; we test our measurement on local data from the SDSS to
show that it provides a robust tracer of environment.
We present the results in $\S$\ref{results}, and discuss the
implications and limitations of this work in \S\ref{discussion}.

\section{Data}\label{survey}
\subsection{Low mass galaxies from ROLES}\label{sec-roles}
In Papers I and II, 
we have constructed a sample of star forming galaxies with stellar mass between $10^{8.5}$ and $10^{10} M_{\sun}$ at $z\sim 1$
using the  [OII]$\lambda$3727 emission line luminosity to derive the SFR.
For a single galaxy [OII] is not an ideal SFR tracer, but it has been
established that, on average, it traces the SFR derived from dust-corrected $H_{\alpha}$ luminosities \citep[e.g.,][]{2001ApJ...551..825J,2004AJ....127.2002K,2010MNRAS.405.2594G}.
With reasonable assumptions about dust and metallicity, we can statistically compare [OII] selected sample in different regimes of stellar mass and environments.

The observations were conducted using the LDSS3 on the 6.5m Magellan at Las Campanas, in nod\& shuffle \citep{2001PASP..113..197G} mode.
The full design of the spectroscopic survey and details of the data reduction and methodology used are given in Paper II, and we briefly recap the most significant points here. We use a red blazed 300 l/mm grism with an average dispersion of $\sim$2.7\AA~per pixel. Our custom KG750 filter has a 
7040-8010\AA~bandpass, and hence allows us to observe
[OII]$\lambda$3727 over $z$=0.889-1.149. The 8 arcmin LDSS3 field of
view typically allows us to allocate almost $\sim 200$ non-overlapping
3 arcsec long slits in this configuration using nod \& shuffle mode.

We observed six masks in each of three pointings; two  within
the GOODS-CDFS, and a third in the FIRES MS1054 field.  These fields
were chosen because they have very deep $K-$band data and publicly
available photometric-redshift catalogues. 
Our GOODS-CDFS parent catalog was kindly provided by \cite{2004ApJ...600L.167M}, and the MS1054 catalog was obtained from the FIRES team \citep{2006AJ....131.1891F}. 
We prioritized galaxies with photometric-redshifts which indicate the
greatest probability that  the [OII] line lies in our redshift range,   A full discussion of the selection and completeness can be found in Paper II, and we revisit the completeness for our purposes in Section~\ref{weight}. 

Following Paper~II, we select emission line objects with greater than 4$\sigma$ significance, and with a high likelihood that the line is [OII] rather than something else, based on the photometric redshift probability.  This leaves us with 311 [OII] galaxies with spectroscopic redshifts from the CDFS and FIRES patches.  Although both fields are statistically equivalent, the smaller FIRES field 
only contributes 75 galaxies to the whole sample.  Moreover, our
analysis will need to
combine our spectroscopic sample with the publicly available CDFS
redshift  data for brighter, higher mass galaxies to allow us to
accurately measure local environment.    Unfortunately, equivalent data
are not available for the FIRES field in our redshift range. 
For these reasons, we choose to omit data from the FIRES field from the present analysis.

In Paper II, we tested and discussed the derivation of SFR using several different methods.
In this paper, we will convert [OII] luminosity to SFR using the empirical, mass-dependent correction introduced in \citet{2010MNRAS.405.2594G}.
In brief, our SFR is estimated from  the \citet{1998ARA&A..36..189K} relation, converted 
to the \citet{2003ApJ...593..258B} IMF and with a mass-dependent correction that accounts for variations in dust, metallicity or other effects with mass. 
The typical derived SFR is $\sim$1$M_{\sun}/$yr at $0.889 \leq z \leq 1.149$, and our sample is complete to SFR $\sim$ 0.3$M_{\sun}/$yr.
We note that AGN are rare in our sample (Paper IV; Gilbank et al., in prep), as
expected due to their low stellar mass  \citep[e.g.,][]{2003MNRAS.341...54K,2004MNRAS.353..713K}.
This calibration is based on a nearby sample of galaxies from SDSS, and
will account for any systematic effect (e.g. dust or
metallicity) that depends only on stellar mass.
In applying this calibration to our present sample, we make a
fundamental assumption that it is also applicable to
galaxies at $z\sim1$.  Since our main purpose is to compare galaxies of
fixed stellar mass, in different environments, our main results are
insensitive to this assumption.  They are, however, dependent on the
assumption that the dust content and metallicity of a galaxy are
primarily determined by its stellar mass, and not by some other
parameter that correlates with environment.

\subsection{Additional Redshifts}
For the GOODS-CDFS field, the public spectra are obtained from ESO
FORS2 observations (v3.0) of \citet{2005A&A...434...53V,2006A&A...454..423V,2008A&A...478...83V}. 
These targets were selected to have $z_{850} < 26$, with $(i_{775}-z_{850})>0.6$ for the primary targets and $(i_{775}-z_{850}) < 0.6$ for the secondary.
We only include those galaxies with a good spectroscopic flag (as `A' or `B') in our spectroscopic catalog, using a 0.5\arcsec\ matching criterion in position. 
This gives us a total of 516 FORS2 spectra.
We compute the [OII] flux from these spectra, and use the method described in \S~\ref{sec-roles} to derive the SFR. 
We also use redshifts from FIREWORKS \citep{2008ApJ...682..985W}, which
provide 2119 spectroscopic redshifts obtained from various sources in
the literature.  We match these redshifts to our photometric catalogue
using a radius of 0.5\arcsec, and find 1463 of them within our two
LDSS3 pointings, and with
spectroscopic quality confidence $z_{qop}\ge 0.5$.  Unfortunately, we do not have [OII] measurements for these spectra; the redshifts will be useful for measuring the local density in our fields, but will not contribute to the SFR measurement.  We will account for this in our weighting scheme, described below.

\subsection{Final Spectroscopic Catalogue and Stellar Masses \label{catalog}}
\begin{figure}
\includegraphics[width=6cm,angle=90]{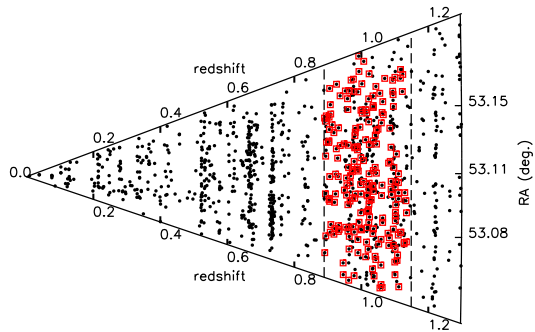}
\caption{The spectroscopic redshift distribution of the 1381 galaxies
  in our CDFS spectroscopic sample.
The sample is generated by combined our LDSS3 spectra with publicly available data as described in the text. The vertical, dashed lines mark our LDSS3 redshift window, $0.889 \leq z \leq 1.149$, and the red squares mark LDSS3 galaxies in this redshift range. A total of 441 spectroscopic galaxies  are within this redshift window.
 \label{fig:zspec}}
\end{figure}
Our final catalog contains 1381 galaxies with multiwavelength optical
and near-infrared photometry, and confident spectroscopic redshifts.  
Among the 441 galaxies at $0.889 \leq z \leq 1.149$, 312 have an available SFR measurement (i.e. from the LDSS3 and FORS2 subsets).
The distribution of spectroscopic redshifts of our final catalog is shown in Fig. \ref{fig:zspec}.
We estimate the stellar mass for individual galaxies using the spectral
energy distribution grid fitting method of \citet{2004Natur.430..181G}, as in Paper II.
We note that the `stellar mass' as defined here refers to the mass currently locked in stars, which is less than the `formed stellar mass', as it excludes recycled 
gas and mass locked in sub-stellar compact remnants. 
The median stellar mass $M_*$ for galaxies with $22.5 \leq K \leq 24.0$ at $0.889 \leq z \leq 1.149$ is log($M_*/M_{\sun}$)$\sim$9.1, and the limit for our sample of star--forming galaxies is  log($M_*/M_{\sun}$)=8.5.
In the rest of this paper we refer to this combined spectroscopic catalog in the CDFS field (LDSS3+FORS2+FIREWORKS) as `ROLES' and those emission galaxies obtained in our own LDSS3 survey as `LDSS3' galaxies.

\subsection{The Photometric-Redshift Catalogs \label{photoz}}
   We use photometric redshifts (hereafter `photo-z') to estimate the completeness, in both SFR and redshifts, of our sample (see \S\ref{weight}).
The public photo-z catalog is obtained from FIREWORKS \citep{2008ApJ...682..985W}, who used a standard template fitting method.
The 1$\sigma$ uncertainty in these  photo-z, compared with spectroscopic redshifts, is 0.074 at z$<$2 over all magnitudes; and
0.061 and 0.114 for galaxies with $K < 22.5$ and $22.5 \leq K \leq 24.0$, respectively.
For the few galaxies in our ROLES catalogue that do not have FIREWORKS
redshifts, we use the photo-z from which our survey was initially
selected, \citet{2004ApJ...600L.167M}, as described in Paper~II.

Since the photo-z are important for our completeness estimates, it is useful to check these using an independent method.  Therefore, we derive our own independent photo-z. 
This allows us to cross-check the stability of our the computation of
local galaxy density, and hence the final results, against quite
different photo-z methods (Appendix~\ref{glsd-test}).
These new photo-z are
calculated using an empirical fitting method modified from \citet{2008AJ....135..809L}. 
This requires a training set, which we construct  using the full spectroscopic catalogue in the CDFS field, including our LDSS3 targets. We modify the method
of  \citet{2008AJ....135..809L} so  that,  instead of using 19 fixed magnitude-color cells to derive the fitting solutions, 
we do the fit using a subset of training-set galaxies which are chosen to have magnitudes and colors closest to a galaxy for which we wish to measure a photo-z. 
This reduces the bias in the fitting solutions when a galaxy is located near the edge of a color-magnitude cell.
During the fitting process, a weight based on the rank of the quadratic sum of magnitudes and colors to the input galaxy is assigned to each training set galaxy in the same dynamical magnitude-color subset; i.e., the training-set galaxies with magnitudes and colors more similar to the input galaxies contribute more weight.
The photo-z PDF and hence uncertainty for each galaxy is computed by simulating galaxy photometry uncertainties with Gaussian distributions, and bootstrapping the training-set galaxies in a magnitude-color cell.
Our modified photo-z method has been tested in
\citet{2010arXiv1008.0658H},  
and has shown robust results in both tests using the mock data and observed GOODS-N data.
The 1$\sigma$ dispersions are 0.040 and 0.114 using the empirical photo-z for galaxies with $K < 22.5$ and $22.5 \leq K \leq 24.0$, respectively.

All the science analyses in this paper have been done using both the public FIREWORKS and our empirical photo-z.  We find that the main difference is in the completeness measures (see \S\ref{weight}),  but our conclusions are insensitive to which set is used.  
\begin{figure*}
\includegraphics[width=8cm,angle=90]{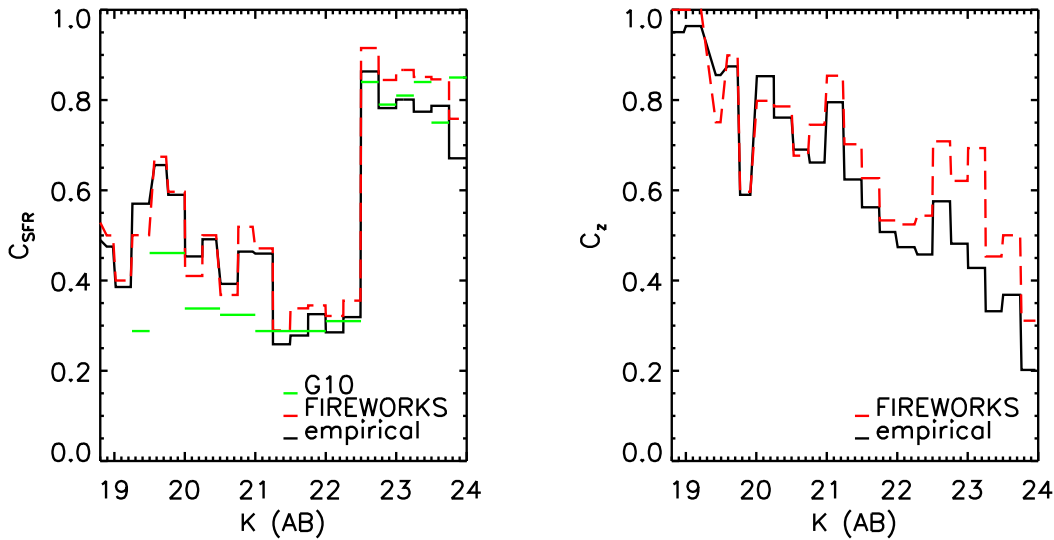}
\caption{Our survey completeness as a function of apparent $K$(AB) in a
  binsize of $\Delta K$=0.25 in our survey. The left panel shows
  $C_{SFR}$, which represents the fraction of galaxies for which we
  have available star formation rate measurements.  The right panel
  shows $C_{z}$, which is our redshift completeness.  These estimates
  rely on the PDF of our photo-z estimators.  The (red) dashed curves
  are computed using the public FIREWORKS photo-z catalogs, while the
  solid lines are derived using our empirical photo-z. The use of
  different photo-z largely affects $C_z$ at $22.5 \leq K \leq 24$. The
  $C_{SFR}$ from Paper II is overplotted as thin green bars (labelled G10). 
\label{fig:w}}
\end{figure*}

\subsection{Completeness \label{weight}}
Following Paper II, we derive {\it a posteriori} completeness corrections for our sample selection strategy, using the photo-z probability distribution function (PDF) of each galaxy.  If a galaxy has a reliable spectroscopic redshift, we replace its photo-z PDF with a $\delta$-function located at the spectroscopic redshift. Since the public data compiled by FIREWORKS provide us with redshifts but not SFR, we must define two completeness corrections: $C_Z$ and $C_{SFR}$, respectively. 
The SFR completeness  $C_{SFR}$ is computed in bins of $K$ magnitude, 
with a binsize $\Delta K$=0.25 as,
        \begin{equation}
        C_{SFR}(K) = \frac{\sum_i \int_{0.889}^{1.149} P_i(z|slit) dz}{{\sum_j \int_{0.889}^{1.149} P_j(z) dz}}.
        \end{equation}

The denominator is the sum of the photo-z PDF of all the galaxies in a K magnitude bin over our redshift range $0.889 \leq z \leq 1.149$. 
The numerator is computed as the sum of the photo-z PDF over the same
redshift range for the galaxies in the same $K$ magnitude bin but with
slits allocated from either our LDSS3 data or the FORS2 catalogue.
This implicitly assumes that redshifts are successfully obtained for
every targeted galaxy that has significant SFR and lies within the
specified redshift range.  This is likely a good approximation for our
faint, LDSS3 sample (see below).  Although the redshift success of the FORS2 spectroscopy is only about 70 per cent \citep{2008A&A...478...83V}, we assume here that redshift failures are either absorption--line galaxies, or galaxies with $z>1.4$, for which the [OII] line would not be accessible.  
A plot of $C_{SFR}(K)$ as a function of magnitude is shown in
Fig. \ref{fig:w}. The values are quite high ($\sim $0.8 in the range
$22.5<K<24$) as we observed nearly all the faint targets in our
redshift range.  For the LDSS3 subsample, this $C_{SFR}(K)$ is analogous to what Paper II terms `$w_k$', which was used to calculate SFR densities (equation 15 of Paper II). 
We compare our completeness estimates with $w_k$ directly in
Fig. \ref{fig:w}, and we find that the two completeness estimates are
similar, but not identical.  The main difference is due to the fact that $w_z$ in Paper~II also accounted for the 70\% redshift success rate of the FORS2 spectra.  This is the correct approach if the completeness is independent of SFR (i.e. emission line presence) and redshift, which seems unlikely.  In any case, this makes little difference for any of our results.

When computing local densities, we can make use of all the spectroscopy, including those public redshifts for which [OII] is unavailable to us.  In this case we consider a redshift completeness
$C_z$, defined as
        \begin{equation}
        C_{z}(K) = \frac{\sum_i \int_{0.889}^{1.149} P_i(z_{spec}) dz}{{\sum_j \int_{0.889}^{1.149} P_j(z) dz}}
        = \frac{N(z_{spec}=0.889-1.149)}{{\sum_j \int_{0.889}^{1.149} P_j(z) dz}}, 
        \end{equation}
where now the numerator is a sum over all galaxies with spectroscopic redshifts.
Because we assume a $\delta$-function redshift PDF for galaxies with spectroscopic redshifts, this is equivalent to the number of galaxies with spectro-z within our redshift range.
We show $C_z$ in Figure~\ref{fig:w}.  Note
that $C_z$ is larger than $C_{SFR}$ at $K < 22.5$, because of the inclusion of public redshifts in FIREWORKS for which SFR measurements are unavailable.

Figure \ref{fig:w} shows that at the faintest magnitudes, where we have
contributed new LDSS3 spectroscopy, the completeness $C_{SFR}\sim 0.8$
is quite high, reflecting the high sampling completeness of our survey
(Paper~II).   The fact that the redshift completeness $C_z$ is also quite high ($\sim0.6$) at these magnitudes, indicates that the majority of these galaxies have ongoing star formation, for we only obtain redshifts for emission--line galaxies at these magnitudes.

We have checked that galaxies with spectroscopic redshifts are distributed uniformly over our field (see Figure \ref{fig:tmap}); thus we do not need to apply a spatial completeness correction.

\subsubsection{Sensitivity to Photometric Redshift Estimates}
We repeat the calculations above, using the PDFs obtained from our empirical photo-z measurements, described in \S~\ref{photoz}. We compare the resulting
 $C_{SFR}$ and $C_z$, as a function of $K$, with our default measurements (described above) in Fig. \ref{fig:w}.
We find that the empirical photo-z tends to return lower $C_{SFR}$ and $C_z$, especially at 22.5$\leq K \leq$24. 
The average $C_{z}$ are [0.62, 0.43] for [22.5$< K$,22.5$\leq K \leq$24] using the empirical photo-z, and are [0.68, 0.62] using the public photo-z.
The average $C_{SFR}$ are [0.47, 0.77] and [0.41, 0.84] over the same magnitude ranges for using the empirical and public photo-z, respectively.
We note that none of our main results are sensitive to which completeness we choose, and we show this explicitly where relevant.  

\subsection{The SDSS comparison sample \label{sdss}}
   The Sloan Digital Sky Survey \citep[SDSS;][]{2000AJ....120.1579Y} is a project that has imaged 10$^4$ deg$^2$ in \it u'g'r'i'z' \rm passbands, and obtained spectra of 10$^6$ objects.
Our sample is selected from Stripe 82, which is part of the SDSS
`legacy survey', and covers an area of 275 deg$^2$
Our catalog is generated using the method described in \citet{2010MNRAS.405.2594G}, 
and contains 47855 objects with $r' < 19.5$ at $0.032 \leq z \leq 0.2$.
The lower redshift boundary is set by the availability of the [OII]$\lambda$3727 line.
Each galaxy is assigned a completeness weight which is calculated using a method similar to that described in \citet{2005MNRAS.358..441B}.
The targeting completeness is nearly 100\% at $r' < 17.8$ and  drops sharply at fainter magnitudes. 
We also match the objects in our sample to those in the SDSS DR4\footnote{http://www.mpa-garching.mpg.de/SDSS/DR4/} to obtain stellar mass measurements, described in \citet{2003MNRAS.341...33K}.
For those galaxies in DR7 but not in DR4, 
we derive their stellar mass use a color-based estimate of the mass-to-light ratio \citep[e.g.,][]{2003ApJS..149..289B}, fit empirically against the \citet{2003MNRAS.341...33K} mass.
The stellar mass obtained in this manner has 1$\sigma$ dispersion of 0.21 dex compared to the value of \citet{2003MNRAS.341...33K}.

Since we intend to use this SDSS sample to provide fiducial comparisons against our $z\sim1$ ROLES sample, 
we limit the catalogue in stellar mass to 
log($M_*/M_{\sun})>8.5$; for star--forming galaxies our sample is complete
for $z<0.05$.

\citet{2010MNRAS.405.2594G} demonstrated that Stripe 82 samples
selected from [OII]--derived SFR  are incomplete at high stellar mass. 
This is primarily because the detection of emission lines is sensitive to the contrast with the continuum; for the brightest galaxies, this imposes a selection limit in SFR which is not low enough to uncover all the star--forming galaxies at that magnitude.  
This problem is mitigated by using SFR derived from Balmer--decremented H$\alpha$ measurements, since this is a stronger emission line \citep{2010MNRAS.405.2594G} .  We will therefore adopt this estimate of SFR for the SDSS sample.
This still allows a fair comparison with our $z\sim 1$ data, since  the [OII]-derived SFR used there are empirically corrected to agree, on average, with the $H\alpha$-derived SFR. 

\section{Local Galaxy Density Measurements}\label{glsd}
   We will adopt a local galaxy density estimator, $\rho_n$, as a tracer of galaxy environment.
This is an $n^{th}$  nearest-neighbor method,  counting galaxies with spectroscopic redshifts 
brighter than a specified luminosity.  Our approach
is to use the brighter galaxies to define the environment, as this
sample is complete for both star-forming and passive galaxies, once
weighted by $1/C_z$.   
Accordingly we consider galaxies with $M_K\leq-$21.0 ($\sim M^*_K+2$)
from the spectroscopic sample to define the environment.  In section
Appendix~\ref{glsd-test} we will consider the
sensitivity of our results to this choice of density estimator.

\begin{figure}
\includegraphics[width=6cm,angle=-90]{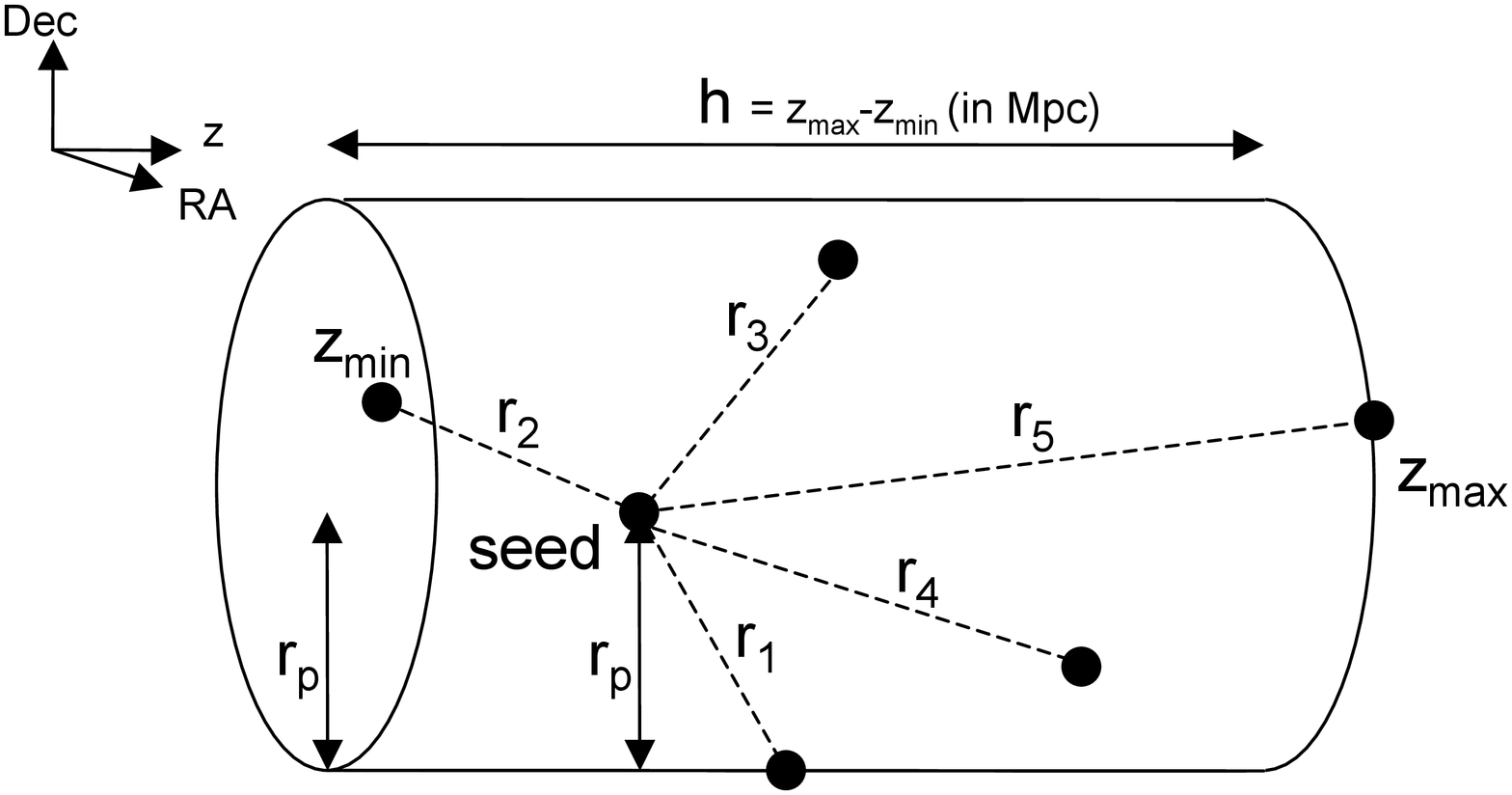}
\caption{An illustration of a cylinder defined by the first five nearest neighbors of a target galaxy, for the purpose of computing local galaxy density. $r_i$ with $i$=1,2,3,...,$n$ are the 3D distances to each neighbor. The radius of the cylinder $r_p$ is defined by the maximum projected distance of all these five nearest neighbors, while the height of the cylinder $h$ is determined by the maximum redshift difference of the same galaxies.
\label{fig:cylinder}}
\end{figure}
\begin{figure}
\includegraphics[width=8cm,angle=90]{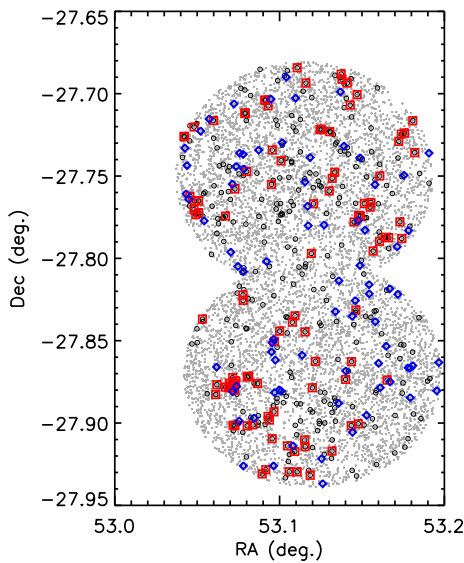}
\caption{The spatial distribution of our spectroscopic sample in the CDFS. 
The gray dots are all galaxies in the photometric catalog, which demonstrates the geometry of our two-pointing LDSS3 configuration. Galaxies at $0.889 \leq z_{spec} < 1.149$ are plotted in black open circles. Those located in the highest and lowest 20\% $\rho_5$ regions are marked by red squares and blue diamonds, respectively.
\label{fig:tmap}}
\end{figure}
\begin{figure*}
\includegraphics[angle=90,width=18cm]{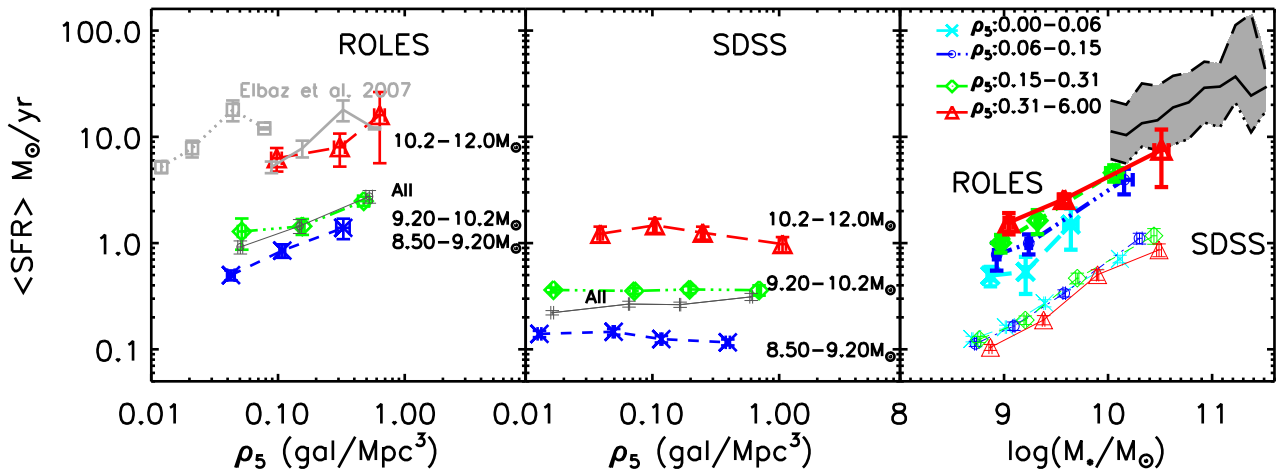}
\caption{{\it Left} and {\it Middle}:  The median SFR is shown as a function of $\rho_5$ in each $M_*$ bin, as indicated in the plot. The trend from the full sample, averaged over all masses, is overplotted as solid gray curve. 
The trend from Elbaz et al. (2007) is overplotted as the gray dotted
curve in the ROLES panel, but shown as the gray solid one after
multiplying their densities by an arbitrary factor of 7.4.  Note their
sample includes galaxies with $M>10^{10}M_\odot$, so is most fairly
compared with our red line.
In each $M_*$ bin, the SFR increases with $\rho_5$ (marginally in the
high $M_*$ bin) in the ROLES sample, while there is no significant
trend in the SDSS sample.
\it Right \rm : The median SFR as a function of $M_*$ with $\rho_5$ controlled.
The gray shaded area is the ``main sequence'' with 1$\sigma$ uncertainty in SFR at $z$=0.98 from \citet{2007ApJ...660L..43N}.
The SFR increases with $M_*$ in both ROLES and SDSS samples.
The environmental density dependence is primarily observed for the lowest mass regime at $z\sim 1$; no such dependence is present at $z=0$.
\label{fig:sfr.all}}
\end{figure*}

For every galaxy in our full catalogue, the 3D angular diameter distances to its $M_K\leq-21$ neighbours are computed;
the distance in the $z$-direction is calculated using Eq. 19 in \citet{1999astro.ph..5116H}.
We compute the volume of a cylinder where the radius and height are determined from the first $n$ nearest neighbors, as illustrated in
Fig. \ref{fig:cylinder}.
We use $r_i$ to denote the 3D distance between the $i^{th}$ nearest neighbor ($M_K\leq-21$) and the target.
The radius $r_p$ of the cylinder is the maximum projected distance of $r_i$ for all such neighbours at $r_i \leq r_n$, while
the height $h$ is determined from the maximum difference in $z$ for the same nearest neighbors.
We note that $r_p$ is not necessarily the same as the projected distance of the $n^{th}$ nearest neighbor in 3D,
and the cylinder is not always centered at the target.
Once $r_p$ and $h$ are determined, $\rho_n$~is then computed as
the total galaxy count $N$ within this cylinder, which has volume $V$=$\pi r_p^2 h$.
Here, $N$ is computed by summing up the redshift completeness weights $w_z$ (=1/$C_z$; see \S\ref{weight}) of all spectroscopic galaxies brighter than $M_K= -21.0$ within the cylinder.
That is, 
        \begin{equation}
        \rho_n = \frac{\sum_{i=0}^n w_z(i)}{V}.
        \label{eq:rho_n}
        \end{equation}

In calculating the volume of the cylinder, we correct for the area of its base if $r_p$ extends beyond the survey limits.
To do so, we first generate a random field mimicking our survey coverage,
with a number density of 3$\times10^6$ random points per square degree.
We then derive the area correction factor $k_A$ as the ratio of the total number of the random points actually enclosed by $r_p$ to the total expected number.
The volume of the cylinder is accordingly adjusted as $V=k_A(\pi r_p^2)h$.

In Fig. \ref{fig:tmap} we present the spatial distribution of our spectroscopic sample, where all galaxies at $0.889 \leq z < 1.149$ are plotted as black circles.
Galaxies located within the highest and lowest 20\% $\rho_5$ regimes
are marked by the red squares and blue diamonds, respectively.  We can see that the
galaxies identified as high-density regions are the most clustered,
suggesting that our density estimator is reasonable.  We will test this
further, in the following subsection.  
Finally, we note that our choice of a 3D, nearest-neighbour density
estimator is one of several density estimators that have been used in
the literature.  In
Appendix~\ref{sec:tracer}, we will explore how our conclusions might be
influenced by different choices. 

\section{Results}\label{results}
We intend to study the interplay between stellar mass and local galaxy
density, as they affect the average SFR and sSFR of star--forming
galaxies.  In all the following analysis, star--forming galaxies are
selected from the ROLES $z\sim 1$ sample by requiring a detection of
[OII] emission.  In the $z=0$ comparison Stripe 82 sample, we require
$sSFR>5\times10^{-12}\mbox{yr}^{-1}$, which we find provides a good
separation between the red sequence and blue-cloud population.
All results use weights that are based on the FIREWORKS photo-z; we have checked that our conclusions are unchanged if we repeat the analysis using our empirical photo-z measurements.

\subsection{The correlation between SFR, mass and environment}
In Fig. \ref{fig:sfr.all} we plot the median SFR as a function of $\rho_5$ for both the ROLES and SDSS samples.
First, we consider the average trend, including all star forming
galaxies regardless of their mass; this is shown as the thin, grey, solid curves.  Locally, there is no significant trend with environment, as expected.  At $z=1$, however, we find that the median SFR in our densest regions is a factor $\sim 3$ {\it larger} than in the lowest--density regions.
The SFRs in our ROLES sample are about $\sim$10 times larger than those in the SDSS sample in the high $\rho_5$ regime, but only a factor of $\sim$4 in the low $\rho_5$ regions.

This echoes the recently discovered `reverse' SFR--density relation at $z\sim1$ \citep[e.g.][]{2007A&A...468...33E,2008MNRAS.383.1058C,2009ApJ...700..971I}.
We compare our data directly with those of \citet{2007A&A...468...33E},
by showing their results as the gray dotted curve  in
Fig. \ref{fig:sfr.all}.   In Appendix~\ref{sec:tracer} we show that
their density estimate is not directly comparable to ours, and in
particular is less sensitive to environment for our particular sample.
Based on
the results of this comparison, we have arbitrarily scaled their densities by a factor of $7.4$.
The magnitude of the trend we observe, for comparably massive galaxies $M<10^{10}M_\odot$, is comparable to that of \citet{2007A&A...468...33E}.

We also show our results divided into different stellar mass bins, in the same figure.  Interestingly, the same conclusions hold for all mass bins: while there is little or no trend of SFR with density locally, we see a significant increase at $z=1$.  The trend appears to be strongest for the lowest-mass galaxies, however.

\begin{figure}
\includegraphics[angle=90,width=8cm]{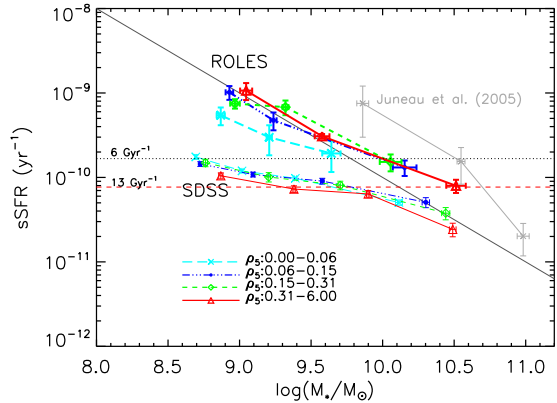}
\caption{We show the sSFR for star forming galaxies as a function of log($M_*$), in different bins of density $\rho_5$, for the ROLES ($z=1$, thick line) and SDSS $z<$0.05 (thin line) samples. 
The solid diagonal line has SFR = 1$M_{\sun}$/yr.
The horizontal dotted and short dashed lines represent $t$=6 and 13 Gyr, corresponding to the age of the Universe at $z\sim 1$ and $z=0$, respectively. 
The data of \citet{2005ApJ...619L.135J} are overplotted as the gray, solid curve for comparison.
The sSFR at fixed $\rho_5$ decreases with log($M_*$) in the both samples.
The lowest sSFR is observed in the lowest $\rho_5$ bin in the ROLES $z\sim 1$ sample.  Locally, the SDSS sample shows little trend with $\rho_5$; if anything, the sSFR is lowest in the highest density environments.
\label{fig:ssfr_combined}}
\end{figure}
The right panel of Fig. \ref{fig:sfr.all} summarizes the SFR--density
relation, now presented as SFR as a function of stellar mass.
Overall, we see the expected increase in SFR with stellar mass, at both
epochs, as has been noted by many others.  For example, we show the
"main sequence" of star--forming galaxies at $0.85<z<1.0$ as measured
by \citet[e.g.][]{2007ApJ...660L..43N} as the gray shaded area in
Fig. \ref{fig:sfr.all}.  Their data, which use 24$\mu m$ luminosities
to estimate SFR, show a similar trend but higher normalization compared
with our data.   We will explore this relation in more detail in a
forthcoming paper (Paper IV; Gilbank et al., in prep).  For now we are interested
in the environmental dependence, and we divide this relation into bins
of $\rho_5$, as indicated on the figure.  The choice of $\rho_5$ bins
in the ROLES sample is determined by requiring a similar number of
galaxies in each $\rho_5$ bin, and we choose the same bins when
considering the SDSS sample.
Both the ROLES and SDSS samples show an increasing median SFR with larger $M_*$ in all environments, and the ROLES sample always has higher SFR than the SDSS. 
Again we observed that the lowest mass galaxies at $z=1$
show a significantly higher SFR in denser environments, while there is little or no trend locally. 

Because massive galaxies possess higher SFR than low-mass galaxies due to their larger integrated gas content, the sSFR provides a better understanding of SFR efficiency.
The sSFR is known to decline with increasing stellar mass both in low- and high-z galaxies, as an outcome of gradual decreasing SFR in galaxies \citep[e.g.,][]{2007ApJ...660L..47N,2007ApJ...660L..43N,2010MNRAS.tmp..588O}. 
Thus, in Figure~\ref{fig:ssfr_combined}, we show sSFR as a function of stellar mass and
$\rho_5$; both ROLES and SDSS show that the sSFR decreases with $M_*$.
The ROLES sample exhibits higher sSFR than the SDSS sample at a fixed $M_*$, and the difference is more significant for lower stellar masses.
We can interpret the sSFR as a ratio of current SFR to the past-average SFR. In the Figure we plot horizontal lines indicating the age of the Universe at $z\sim1$ (gray dotted line) and $z\sim0$ (red short dashes). 
Galaxies below these lines are inefficient in forming stars, i.e. their current SFR is much less than their past average. Galaxies above the lines are more active and would be interpreted as having a current burst of SFR such that it exceeds the past average.
The separation between the `active' and `passive' SFR regimes occurs at
log($M_*/M_{\sun}$)$\sim$10 at $z\sim1$ but at
log($M_*/M_{\sun}$)$\sim$9.6 at $z\sim0$, showing a mass dependence in
the sense that that the contributors of SFR have shifted toward lower $M_*$ as redshift decreases. 
This provide evidence for down-sizing in SFR in the respect of the `cross-over' time between the `active' and `passive' SFR regimes. This accords with previous work such as that of Juneau et al. (2005),
which is overplotted as the gray solid curve, with the SFR recomputed using the mass-dependent correction in Paper II.
However unlike that work, our census is complete at $z\sim 1$ for galaxies with masses below the cross-over mass. 
We explore this further in Paper IV (Gilbank et al. in prep). 

Dividing the samples by $\rho_5$, we find that the evolution of this cross-over time holds in all environments.
Only the lowest-mass star--forming galaxies at $z=1$ (less than $\sim$log($M_*/M_{\sun}$)=9.5) exhibit an increasing sSFR with $\rho_5$.   No such dependence is seen in the local comparison sample; if anything, there is a mild {\it decrease} in the mean sSFR at the highest densities, locally \citep[see also, e.g. ][]{2004MNRAS.348.1355B, 2009ApJ...705L..67P, 2005AJ....130.1482R, 2009MNRAS.398..754B}.  
%
In other words, the lowest-mass star-forming galaxies in high-density regions have the largest decrease in sSFR from $z$=1 to $z\sim0$.
This suggest that the evolution of galaxies with
$M<10^{10}M_{\sun}$ is characterised by a
more rapid decrease of star formation in moderately dense regions.

\subsection{Star formation density}\label{subsec:SFRD}
\begin{figure}
\includegraphics[angle=90,width=8cm]{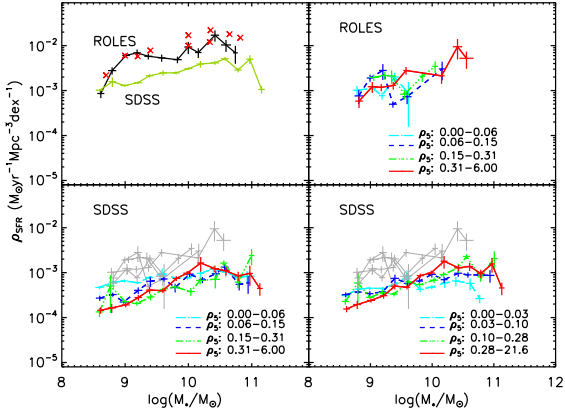}
\caption{
The SFRD is shown as a function of log($M_*/M_{\sun}$), using a bin size of $\Delta$log($M_*/M_{\sun}$)=0.2.
In the top-left panel, galaxies in all $\rho_5$ bins are combined together for the ROLES ($z\sim 1$, black) and SDSS $z<$0.05 (green) samples. 
The data of the $z\sim1$ sample in Paper II are overplotted as red crosses.
The top-right panel shows the same data, divided into bins of different environment, as measured by $\rho_5$.
The results for the SDSS are presented in the bottom two panels.  In the lower-left panel, the $\rho_5$ bins are chosen to exactly match those in ROLES; in the right panel we choose the $\rho_5$ bins to include an equal number of SDSS galaxies in each bin.
The gray curves in each SDSS panel are the ROLES results from the top-right panel, for comparison.
The ROLES sample exhibits a larger SFRD at all stellar masses than the SDSS sample, reflecting the decline of global SFRD.   Little or no environmental dependence is observed at $z=1$, which suggests that the small trend we observe --- for low-mass galaxies to have higher SFR in denser regions --- has little effect on the global SFRD.
\label{fig:sfrd_combined}}
\end{figure}

Paper II presented the dependence of SFR density ($\rho_{SFR}$) on stellar mass in the ROLES survey, finding the same shape
as seen in the local SDSS. In 
Fig. \ref{fig:sfrd_combined} we now extend this to show $\rho_{SFR}$ as a function of $M_*$ in different environments, as measured by $\rho_5$.
The $\rho_{SFR}$ in each $M_*$ bin is computed following Paper II as 
        \begin{displaymath}
        \rho_{SFR} = \sum_i \frac{w_{SFR,i}SFR_i}{V_{max,i}},
        \end{displaymath}
where $w_{SFR,i}=1/C_{SFR,i}$ (see \S\ref{weight}) and $V_{max,i}$ is the maximum volume in which galaxy $i$ could be located and have been found as $K \leq 24$ and $0.889 \leq z \leq 1.149$. 

Including galaxies in all environments, our results in the top-left panel in Fig. \ref{fig:sfrd_combined} are consistent with those of Paper II.  
We next present this same distribution in different $\rho_5$ bins, with results shown also in Fig. \ref{fig:sfrd_combined}.
The trends are noisy, especially for the ROLES sample, due to the small sample size in each $\rho_5$ and log($M_*)$ bin.
Nonetheless, we see that, in the local sample, the lowest-mass galaxies
(log($M_*/M_{\sun})<9.5$) tend to exhibit a decreasing $\rho_{SFR}$ with increasing density. At higher masses, there is little or no dependence on environment.  In our $z=1$ ROLES sample, there is no clear dependence of $\rho_{SFR}$ on environment.  The small increase we see in median SFR toward higher densities appears to have little noticeable impact on the total SFR density.

The evolution we see indicates that environment accelerates quenching of the lowest-mass galaxies with time, but that $\rho_{SFR}$ is governed primarily by the stellar mass of the galaxy.

\section{Discussion and Conclusions} \label{discussion}
We have used our highly sensitive, complete LDSS3 spectroscopy in the CDFS field to probe the effects of environment on galaxy evolution at $z=1$, for galaxies with masses lower than have ever been probed in this way before.   We consider only star--forming galaxies, for which our survey, and $z=0$ comparison sample, are complete for stellar masses $M>10^{8.5}M_\odot$.
Locally, star forming galaxies show little dependence on environment.
Specifically, the correlation between SFR and mass, or between sSFR and
mass, shows no dependence on local density.  The situation is quite
different at $z=1$, where we find SFR and sSFR is actually somewhat
{\it larger} in dense environments.  Thus we confirm the general trend
found by others
\citep[e.g.][]{2007A&A...468...33E,2008ApJ...686..966M,2008MNRAS.383.1058C,2009ApJ...700..971I};
moreover, we find that this trend is stronger for galaxies with masses below those studied in previous work.  This has little effect on the total SFR density however which, at fixed stellar mass, shows no clear trend with environment.

In summary, we find that the decline
in SFR-mass and sSFR-mass relations from $z\sim 1$ to $z \sim 0$ depends on environment, with a faster decline
in richer environments.  Note that, due to the small field size, our
sample does not include the densest environments, such as the cores of
rich clusters.  Recent results from \citet{2010arXiv1007.2642S} suggest
that it is in these intermediate density environments that star
formation may be most enhanced, at $z\sim 1$.

However these effects are small when compared to the intrinsic dependence of SFR and sSFR on mass
itself.  In agreement with other studies, we conclude that mass is the
dominant variable, driving the history of star-formation \citep[e.g.][]{2010arXiv1003.4747P,2010A&A...509A..40I,2010arXiv1007.3841C}.  Environment appears to play a secondary role, albeit one that is more important for lower--mass galaxies.  This may be expected, since  the smaller gas reservoir of less massive galaxies may be more easily affected by environment.
Mechanisms such as galaxy mergers and harassment are are more likely in  denser environments, and these may enhance the star formation activity within the galaxies, resulting in the positive SFR-$\rho_5$ and sSFR-$\rho_5$ trend we see at $z\sim1$.
Since such vigorous star formation will speed up the consumption of cold gas within the galaxies, they would more quickly exhaust their gas supply.  At later times,
the star-formation rates would be then be {\it less} than the counterparts in poorer environments. 

While this work is among the first to spectroscopically probe the
low-mass regime at $z\sim 1$, it suffers from limitations that can be
addressed in future work. The main one is statistics: error bars can be
greatly improved by larger area, future surveys. 
Also ROLES lacks
sufficient volume to probe the densest environments (rich clusters) in
which the most massive galaxies are found; interesting, recent work by
\citet{2010arXiv1007.2642S} suggests that the enhancement in SFR we
find may not appear in such massive clusters.   Wider, or cluster--targetted, surveys in the future will be able to address this. 
Future papers based on ROLES will explore the lower redshift regime in our survey. This will allow evolution to be explored further but also more extreme environments
to be probed. This is possible because 
the FIRES patch contains a massive cluster, MS1054-03, at $z$=0.83, and the CDFS patch exhibits a large-scale filament (and galaxy groups) at $z$=0.73. 

Our work is based on a key assumption, that the mass-dependent conversion from [OII]
to SFR derived in the local universe by \citet{2010MNRAS.405.2594G} can
be applied to $z\sim1$.  As dust plays an important role in deriving
SFR from [OII] flux, further inspection will be made using 24$\mu$m and
radio data in the future (Gilbank et al., in prep). However, in general, low-mass galaxies contain little dust and hence are undetected in those bands. 
Since our key results lie in low-mass galaxies, 
and we are comparing trends at a fixed stellar mass, 
any systematics related only to mass should not affect our results
about the relative star formation rates in different environments.

Finally we note that  our work has focused on using local density estimators
as a simple proxy for environment. An alternative approach worth pursuing would be to use linking algorithms to try and identify actual physical galaxy groups. This may
allow a better connection from observables to theoretical concepts such as halo and sub-halo masses. 

\section{Acknowledgments}\label{sec-akn}

This paper includes data gathered with the 6.5-m Magellan Telescopes located at Las Campanas Observatory, Chile. We thank LCO and the OCIW for the allocation of time to this project as part of the LDSS3 instrument project. 
Australian access to the Magellan Telescopes was also supported through the Major National Research Facilities and the National Collaborative Research Infrastructure Strategy programs of the Australian Federal Government. 
KG and I.H.L. acknowledge financial support from Australian Research Council (ARC) Discovery Project DP0774469. MLB acknowledges support from the province of Ontario in the form of an Early Researcher Award.



\appendix
\section{Tests of Local Galaxy Density }\label{glsd-test}
In this Appendix, we explore the sensitivity of our main results to our choice of density estimator.  
Our measurement of environment is an estimator based on a biased,
incomplete tracer population, and we expect that its correlation with a
fundamental quantity (such as dark matter density field) will be both
biased and noisy.  We will not attempt to directly connect our estimates
directly to more physical parameters like these; however, we will ensure that we are able to distinguish high-density environments from low.  
To do this, we will use the local SDSS sample, which benefits from both large area and high spectroscopic completeness.  
We first apply our computation of $\rho_n$ to the SDSS sample described in \S~\ref{sdss}, and compare with the density estimates of \citet{2006MNRAS.373..469B}.
The latter are computed based on a 2D,  n$^{th}$-nearest-neighbor method, for galaxies
limited to $M_{r'}=-20-Q(z-z_0)$ with Q=1.6 and $z_0$=0.05 from \citet{2003ApJ...592..819B}.
The local galaxy density $\Sigma_5$ is calculated from the projected distance to the 5$^{th}$ nearest neighbor, within a redshift slice of $\Delta cz$=1000km/s centered at each seed galaxy.
By comparison with simulations and models, \citet{2006MNRAS.373..469B} show that 
$\Sigma_5$ correlates well with the smoothed, dark matter overdensity.  We will therefore take this as our best estimate of the "true" density in the SDSS, and in the following subsection we will compare how our methods and sample limitations compare with this.
\begin{figure*}
\includegraphics[width=7cm,angle=90]{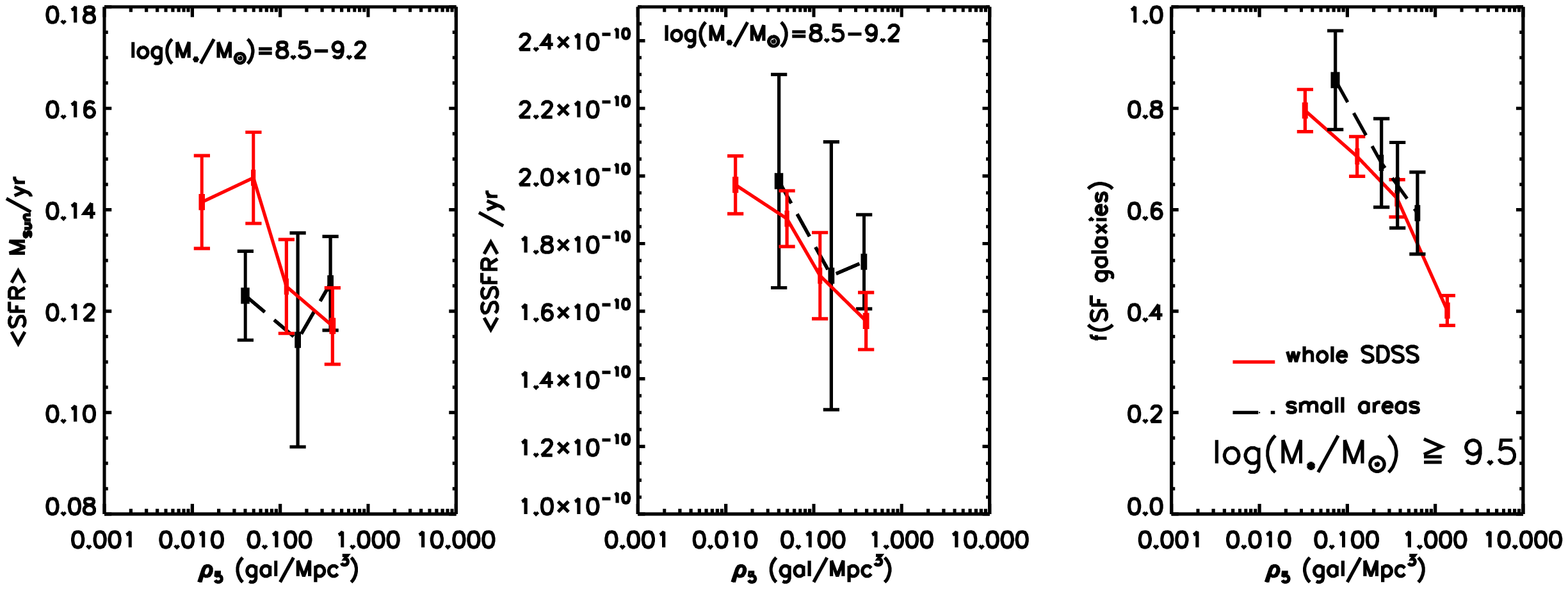}
\caption{To test our measure of environment, we use the SDSS Stripe 82 sample.  We select random regions with a similar physical size to one of our LDSS3 pointings at $z=1$, and measure the local density $\rho_5$.  We then combine those 28 fields that have an average density equal to the average density in our ROLES field, to within 20 per cent.  The
{\it Left} and {\it Middle} panels show the SFR-- and sSFR--$\rho_5$ trends for star-forming galaxies with log($M_*/M_{\sun}$)=8.5-9.2 at $z\leq$0.05 from this experiment, as the black line.  This is compared with the result obtained from the full Stripe 82 sample, shown as the thick red curve.  No strong trend with environment is seen in any case.
The right panel is similar, but for the fraction of star-forming galaxies with  log($M_*/M_{\sun}$)$\geq$9.5.  Here, the well-known trend with environment is recovered, even with small field sizes.  Thus, we conclude that our survey is large enough to detect environmental trends like those present at $z=0$. 
\label{fig:dentest_sdss}}
\end{figure*}

\subsection{Sensitivity to the limited field size}
Likely the biggest limitation of our $z=1$ sample, is the field size
\citep[see discussion in ][]{2010arXiv1007.1967C}.  Our LDSS3 spectroscopic coverage of CDFS is comprised of two adjacent, circular fields with diameter 8.3\arcmin, corresponding to $\sim$ 4.1 Mpc at $z\sim1$.    Thus, we do not expect a large dynamic range of environment within our sample.  Moreover, our density estimator $\rho_n$ effectively smooths this density field over fairly large scales.
The choice of using smaller $n$ results in higher resolution, but a noisier density estimate; on the other hand,
a larger $n$ reduces the difference between the extreme high and low density regions and smears out any trends which depend on local galaxy density.  Edge corrections also become more important for larger choices of $n$.  
With our default choice of $n$=5, the median $r_p$ is $\sim$2.6 Mpc
($\sim$5.4\arcmin at $z$=1) and $h$ is $\sim$3.8 Mpc ($\Delta z
\sim$0.003).  The spatial component is thus actually larger the size of
our field for median densities; thus, most of the sensitivity comes from the
redshift dimension.  Nonetheless, as we will show, it is still possible to distinguish high- and low-density environments.

We carry out the test by randomly choosing 500 circular regions from
the SDSS Stripe 82 sample, each with a diameter of 4.1 Mpc at $z$=0.05.
Thus, the field size is matched to the same physical area covered by a
single LDSS3 pointing at $z=1$.  
In each field we compute $\rho_n$ with $n$=[3,5,7]. 
Forty of these 500 circular regions are rejected because their mean $\rho_n$ are significantly larger than that in ROLES.  The computed $\rho_n$ for these SDSS galaxies have a 1$\sigma$ dispersion of [0.46, 0.22, 0.08], and the median $r_p$ are [1.39,2.16,2.8] Mpc for $n$=[3,5,7].   We compare these $\rho_n$, computed within the limited field size, with our best estimate of the "true" density field, $\Sigma_5$.  We find a 1$\sigma$ dispersion between $\rho_n$ and $\Sigma_5$ are [0.67, 0.49, 0.47] dex for $n$=[3,5,7].  Thus, for our default choice of $n=5$, we can hope to distinguish environments separated by $\sim 1$ dex with $2\sigma$ confidence.   We will show that our $z=1$ sample spans a factor of $\sim 100$ in $\rho_5$; thus we anticipate that differences between the highest and lowest-density environments in that sample will be meaningful.

We proceed to check this further, by comparing our measurement of $\rho_5$ in SDSS with quantities that have known environmental trends.  In particular, we will consider the SFR and the specific SFR (sSFR), the relevant quantities for our $z=1$ analysis. To improve the statistics, we stack the 28 circular regions that have an average density within 20\% of each other, and comparable to the average density in our $z=1$ sample.  This ensures that we have enough galaxies to reliably measure average galaxy properties, and that the dynamic range of density probed will be comparable to what we can expect from our survey.  Note the stacking does not reduce the noise, or potential bias, in our density esimate, since these estimates are still made on the individual fields.

We show the average SFR and sSFR as a function of $\rho_n$ with $n$=5 in the stacked regions in Fig. \ref{fig:dentest_sdss}.  We only include star--forming galaxies here, by selecting those with $sSFR>5\times10^{-12}$yr$^{-1}$.  This effectively removes passive, red-sequence galaxies from the sample.  We also limit this test to galaxies with  log($M_*/M_{\sun}$)=8.5-9.2, which is the mass range of primary interest in this paper.
In both cases, there is little trend with environment; this result is
expected since we are only considering the star--forming population.
This demonstrates that systematics in our density estimator do not
induce an artificial correlation in sSFR.  

The next test is to consider the fraction of star--forming galaxies
($sSFR>5\times10^{-12} \mbox{yr}^{-1}$) in the SDSS, since this is known to correlate strongly with density.  Here, we need to consider a sample limited at a higher stellar mass, log($M_*/M_{\sun}$)$\geq$9.5, because the passive galaxy population becomes incomplete below that limit.  We show this fraction as a function of $\rho_5$, limited to small fields as before, as the black dashed curve in 
the right panel of Fig. \ref{fig:dentest_sdss}.  Again, this is compared with the trend obtained using the full survey area, shown as the red curve.  We see that a strong environmental dependence is still observed from the limited field areas, with surprisingly little evidence for any bias or smoothing of the trend.  
Therefore, we conclude that despite the small field, our choice of
density measure is robust and able to differentiate truly different
environments, as also suggested in Figure~\ref{fig:tmap}.   

\begin{figure}
\includegraphics[width=8cm]{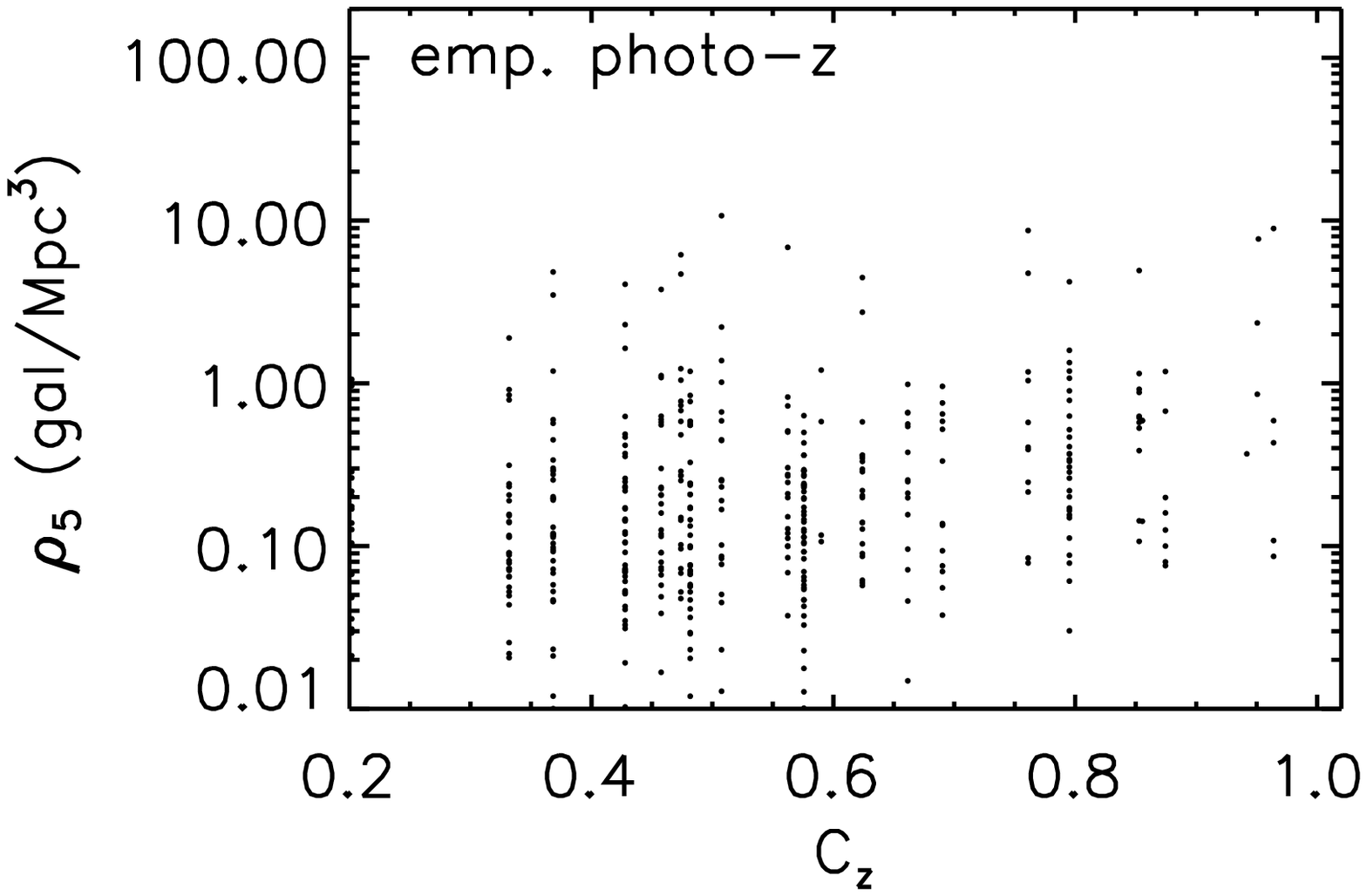}
\includegraphics[width=8cm]{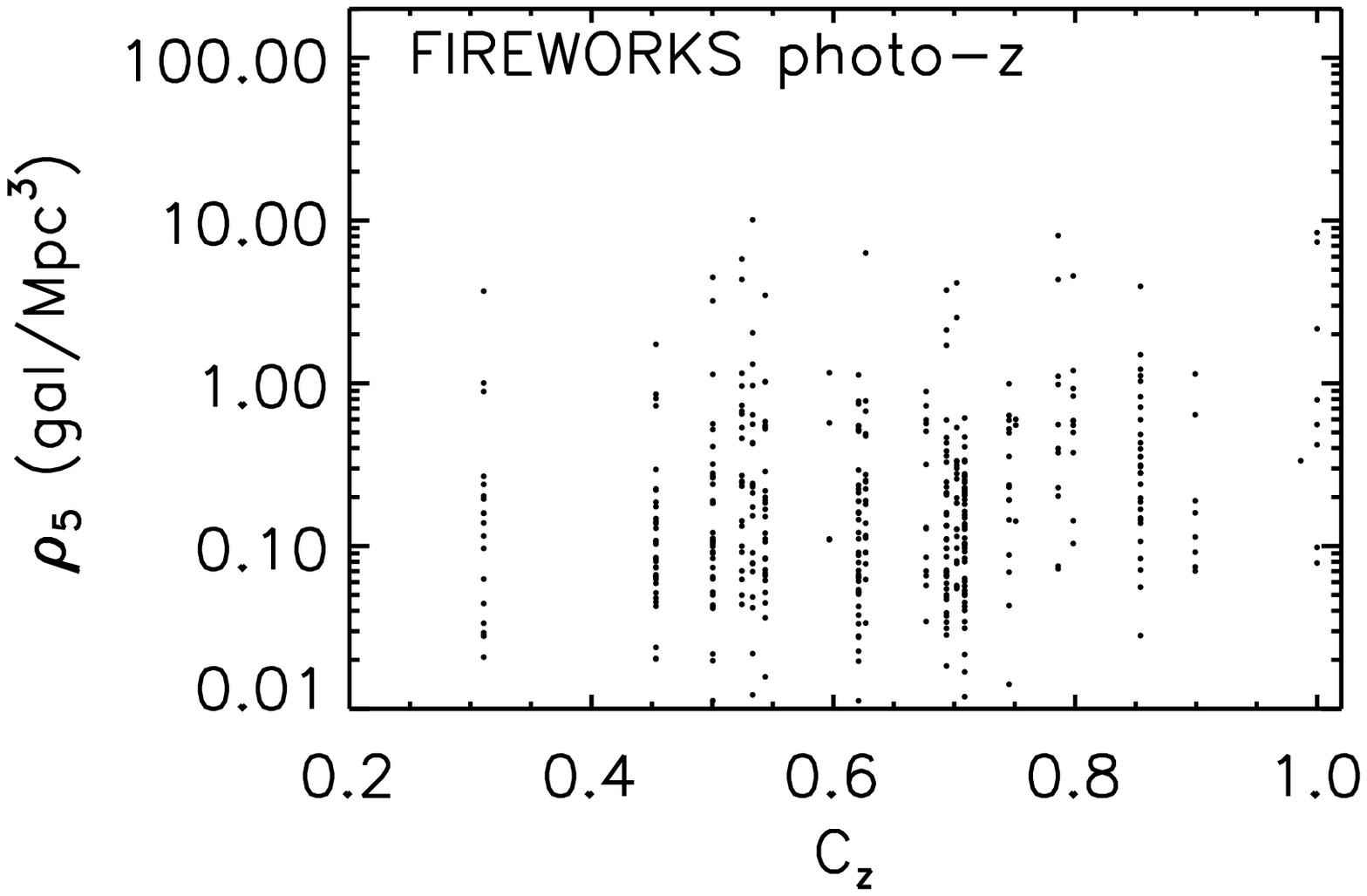}
\caption{The local galaxy density $\rho_5$ is shown as a function of redshift completeness $C_z$ for galaxies at $0.889 \leq z \leq 1.149$.  Completeness are estimated using either our empirical (\it top \rm) photo-z, or the public FIREWORKS (\it bottom \rm) photo-z.
We find that $\rho_5$ is not a strong function of $C_z$, and is not sensitive to our photo-z estimate.
\label{fig:glsd_wz}}
\end{figure}

\subsection{Sensitivity to incompleteness}
Next, we consider the effect of our completeness weights, and their sensitivity to the photo-z estimates (which is the only place the photo-z play a role in our analysis).  
We examine how the choice of photo-z method might affect our computation of $\rho_n$ in Fig. \ref{fig:glsd_wz}, where we choose $n=5$, 
and plot $\rho_5$ as a function of redshift completeness ($C_z$).
Regardless of which photo-z method is used, $\rho_n$ is not a strong function of redshift completeness.  The {\it rms} dispersion between the two measurements of $\rho_5$, on a galaxy-by-galaxy basis, is $\sim 0.025$ dex.  
Thus we find that $\rho_5$ is fairly insensitive to the completeness correction, and thus to which photo-z method is used.

We now consider whether $\rho_5$ itself is biased in the presence of spectroscopic incompleteness.
The exercise is conducted by applying a spectroscopic incompleteness to our SDSS Stripe 82 sample, which is similar to the $C_z$ in ROLES.
To do so, we take the $C_z$ as a function of $K$ in Fig.\ref{fig:w}
(the FIREWORKS curve), and then convert it into a function of $r'$
based on the magnitude relative to $M^*$.   
We use $m_K^*$=20.34 and $m_{r'}^*$=17.26. This produces a 
$C_z$ that starts to drop from 1 at $r'$=16 to $C_z\sim$0.52 at $r'$=19.0.
Galaxies are then removed randomly in each magnitude in to satisfy the `expected' $C_z$ in that magnitude bin. 
The original completeness in the SDSS sample is also accounted for during this process.
The re-computed $\rho_5$ correlates well with the original (complete) $\rho_5$ estimate, with a 1$\sigma$ dispersion of only $\sim$0.02.  We find no evidence for a bias at high or low densities from this experiment.

>From these tests we conclude that our $\rho_5$ measure is robust, and a reasonable proxy for galaxy environment in our sample.

\subsection{Sensitivity to choice of environment tracer}\label{sec:tracer}
\begin{figure}
\includegraphics[angle=90,width=8cm]{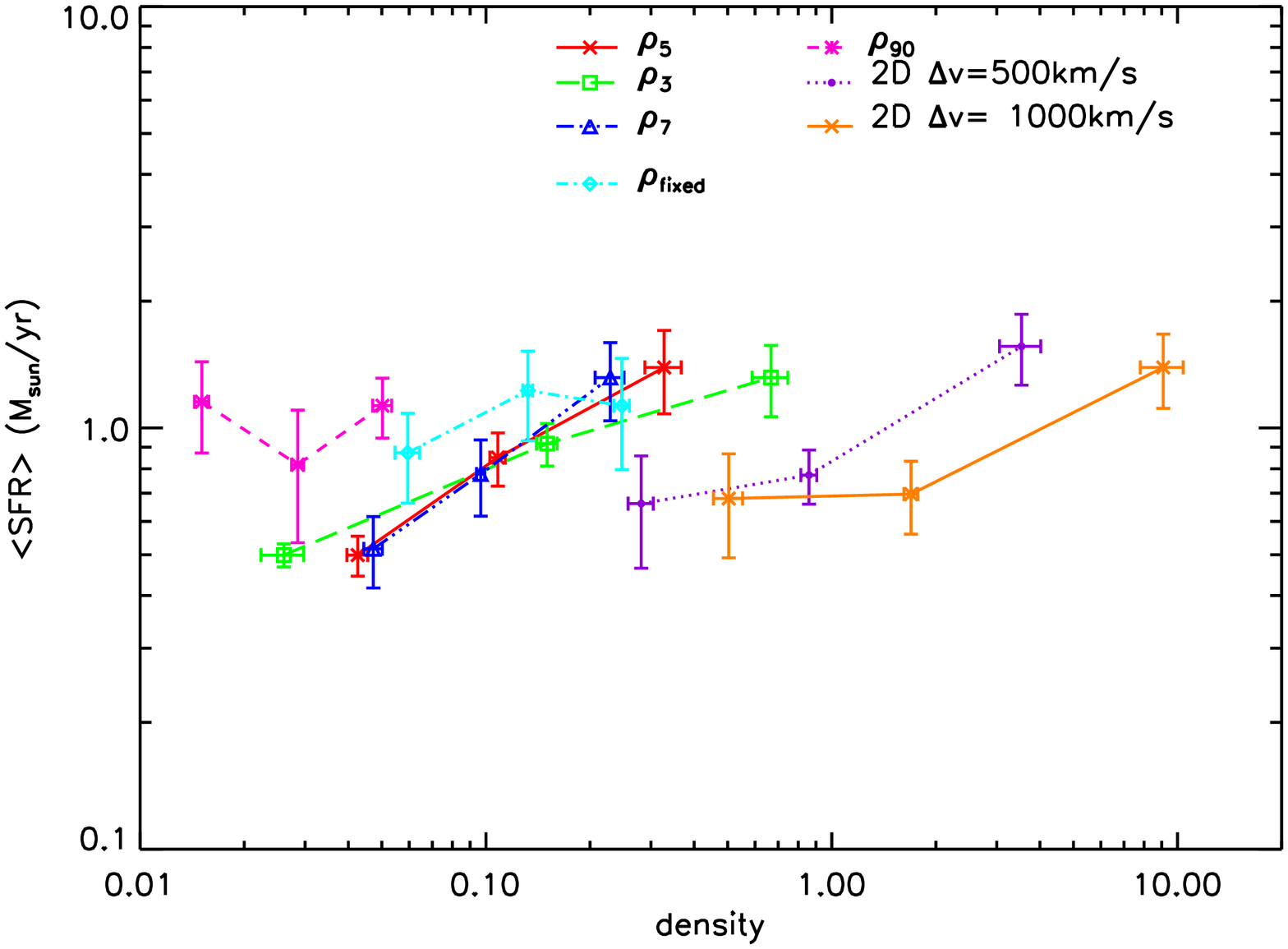}
\includegraphics[angle=90,width=8cm]{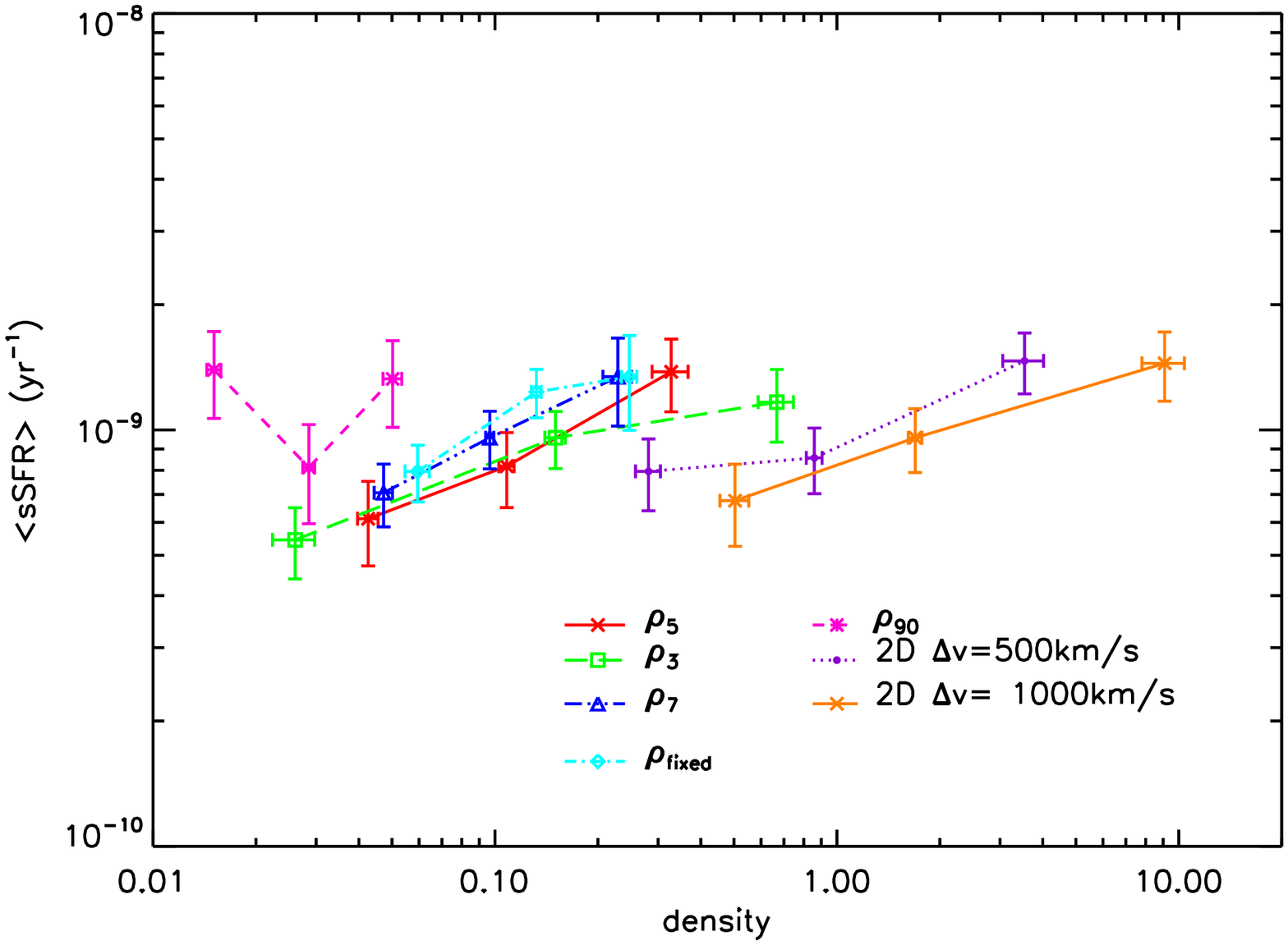}
\caption{The SFR-- and sSFR--density trends are shown, for the ROLES
  sample at $z=1$. The densities are computed with the various methods
  described in this Appendix, as indicated in the plots. Only galaxies with log($M_*/M_{\sun}$)=8.5-9.2 are used in plotting the SFR-- and sSFR--density relation. The conclusions that both SFR and sSFR increases with density remain for most choices of densities.  The only exception is $\rho_{90}$, which is chosen to match the fixed--volume density estimate of Elbaz et al. (2007); our spectroscopic sample of environment tracers is too sparse to yield a useful environment measurement with that definition.  
\label{fig:dentest_roles}}
\end{figure}

Our choice of a 3D, nearest-neighbour density
estimator is one of several density estimators that have been used in
the literature.  In this subsection, we will explore how are results
would be affected if we adopted a different method.  We
thus consider the SFR and sSFR of our lowest--mass galaxies in the $z=1$ sample,  $8.5<$log($M_*/M_{\sun}$)$<9.2$.  Figure \ref{fig:dentest_roles} shows how this depends on density for the variety of possible choices we will explore below, compared with our default choice of $\rho_5$.

First, we consider the sensitivity to the smoothing length by considering the effect of different choices of $n$, by showing results for $n=3,5$ and $7$ in Figure \ref{fig:dentest_roles}.  In all three cases we observe that the average sSFR and sSFR of the star--forming population {\it increases} with increasing density.  This, our main result, is not sensitive to the choice of $n$.

Next, we consider the effect of using a cylinder of fixed volume to compute density.  This has the advantage that the smoothing scale does not depend on density, but has the disadvantage that the noise in the estimator does depend on density.   We choose a fixed 
cylinder with $r_p$=2.6 Mpc and $h$=3.8 Mpc, and re-compute the density.
The density is then computed as the sum of $w_z$ of galaxies within the cylinder divided by the cylinder volume, including a correction for edge effects.
We denote this density as $\rho_{fixed}$, and show the results in
Fig. \ref{fig:dentest_roles}.  Again, the trend of increasing sSFR and
with density is unchanged; however the correlation between SFR and
density has become insignificant.  This is likely due to the fact that
a fixed volume results in a large uncertainty in the lowest--density
regions, as can be seen by the larger error bars of this measure in Figure~\ref{fig:dentest_roles}.

Elbaz et al. (2007) compute a density within a comoving box, with dimension of 1.5$\times$1.5$\times$40 Mpc (= 90 Mpc$^3$).
Since we compare our SFR--density relation to theirs in \S\ref{results}, we calculate this density for our sample, and label it $\rho_{90}$ in Fig. \ref{fig:dentest_roles}.  The resulting densities are approximately an order of magnitude smaller than $\rho_5$, and no correlation with SFR or sSFR is observed.  This is because we have too few spectroscopic density tracers to sufficiently populate the fixed box size.  Thus it is not feasible for us to use this definition for our sample. 

An alternative option is to use a 2D, projected density.  This is widely used in the literature, and we adopt the method described in \citet{2006MNRAS.373..469B}. 
We explore the effect of making two redshift slices of $\Delta$$\frac{cz}{1+z}$=500 and 1000 km/s. 
Within each redshift slice centered, at a target, galaxy, the local galaxy density is computed based on the projected distance to the nearest 5$^{th}$ neighbor.
The SFR-- and sSFR--density relations using these projected densities are also presented in Fig. \ref{fig:dentest_roles}.  The same trend we detect in $\rho_5$ is apparent using these projected estimators.   
\end{document}